\documentstyle[11pt,aaspp4]{article}

\font\caps=cmcsc10 scaled 1200
\def\etal {et~al.}
\def\radm2{radians m$^{-2}$}
\def\deg{{$^\circ$}}

\slugcomment{Accepted to the Astrophysical Journal} 

\lefthead{TAYLOR}
\righthead{MAGNETIC FIELDS IN QUASAR CORES II}

\begin{document}

\title{~~\\ ~~\\ Magnetic Fields in Quasar Cores II}

\author{Gregory B. Taylor}

\affil{National Radio Astronomy Observatory, P.O. Box O, Socorro, NM, 87801, USA}
\authoremail{gtaylor@nrao.edu}

\begin{abstract}

  Multi-frequency polarimetry with the Very Long Baseline Array (VLBA)
  telescope has revealed absolute Faraday Rotation Measures (RMs) in
  excess of 1000 rad m$^{-2}$ in the central regions of 7 out of 8
  strong quasars studied (e.g., 3C\,273, 3C\,279, 3C\,395).
  Beyond a projected distance of
  $\sim$20 pc, however, the jets are found to
  have $|$RM$|$ $<$ 100 rad m$^{-2}$.  Such sharp RM gradients cannot
  be produced by cluster or galactic-scale magnetic fields, but rather
  must be the result of magnetic fields organized over the central
  1--100 pc.  The RMs of the sources studied to date and the
  polarization properties of BL Lacs, quasars and galaxies are shown
  to be consistent so far with the predictions of unified schemes.

  The direct detection of high RMs in these quasar cores can explain
  the low fractional core polarizations usually observed in quasars at
  centimeter wavelengths as the result of irregularities in the
  Faraday screen on scales smaller than the telescope beam.
  Variability in the RM of the core is reported for 3C\,279 between
  observations taken 1.5 years apart, indicating that the Faraday
  screen changes on that timescale, or that the projected superluminal
  motion of the inner jet components samples a new location in the
  screen with time.  Either way, these changes in the Faraday screen
  may explain the dramatic variability in core polarization properties
  displayed by quasars.

\end{abstract}


\vspace{2cm}

\section{Introduction}

At sufficiently high angular resolution ($\sim$1 mas), the inner
jets of some quasars have been found to exhibit Faraday Rotation
Measures in excess of 1000 \radm2\ (Udomprasert \etal\ 1997; Cotton
\etal\ 1997; Taylor 1998).  This RM is most likely produced by
organized magnetic fields and ionized gas close to the center of
activity.  Beyond a projected distance from the core of 20 pc, the RMs in the
jet fall below 100 \radm2 and are thus consistent with a purely
galactic origin.

In \S2 a sample of 40 strong, compact galaxies, quasars and BL
Lacertae objects is defined.  Four of these sources were reported on
in Taylor 1998 (Paper I).  In \S3 I describe observations of four more
quasars from this sample, and the results are presented in \S4.  In
\S5 the RM observations are discussed in the context of unified schemes.

I assume H$_0 = 50$ km s$^{-1}$ Mpc$^{-1}$ and q$_0$=0.5 throughout.

\section{Sample Selection}

The sample of target sources was drawn from the 15 GHz VLBA survey of
Kellermann \etal (1998).  This is a sample of 132 strong, compact AGN
drawn from the complete 5 GHz 1 Jansky catalogue of K\"uhr \etal\
(1981) as supplemented by Stickel, Meisenheimer \& K\"uhr (1994).  To
reduce the sample size to a tractable level and facilitate scheduling,
an initial flux limit of $S_{15}$ $>$ 2 Jy, where $S_{15}$ is the
total 15 GHz flux measured by Kellermann \etal, and declination limit
of $\delta > -$10\deg\ was imposed.  In addition, the somewhat weaker
quasars 3C\,380 and 3C\,395  were also
observed.  The final sample consists of 40 objects and is presented in
Table 1.  In the sample there are 25 quasars, 5 galaxies, 8 
BL Lacertae objects, and 2 as yet unidentified objects.  Redshifts
are available for all but 3 sources.  

\section{Observations}

\subsection{Data Reduction Procedures}

The observations, performed on 1998 Aug.\ 3 (1998.58), were carried
out at multiple frequencies (or IFs) between 4.6 and 15.2 GHz (see Table 2)
using the 10 element VLBA.\footnote{The National Radio Astronomy
Observatory is operated by Associated Universities, Inc., under
cooperative agreement with the National Science Foundation} Both right
and left circular polarizations were recorded using 1 bit sampling
across a bandwidth of 8 MHz.  The VLBA correlator produced 16
frequency channels across each IF during every 2 second integration.

Amplitude calibration for each antenna was derived from measurements
of the antenna gain and system temperatures during each run.  Global
fringe fitting was performed using the AIPS task FRING, an
implementation of the Schwab \& Cotton (1983) algorithm.  The fringe
fitting was performed on each IF and polarization independently using
a solution interval of 2 minutes, and a point source model was
assumed. Next, a short segment of the cross hand data from the
strongly polarized calibrator 3C\,279 (1253{\tt -}055) was fringe
fitted in order to determine the right-left delay difference, and the
correction obtained was applied to the rest of the data.  Once delay
and rate solutions were applied the data were averaged in frequency.
The data from all sources were edited and averaged over 20 second
intervals using {\caps Difmap} (Shepherd, Pearson \& Taylor 1994;
Shepherd 1997) and then were subsequently self-calibrated using AIPS
and {\caps Difmap} in combination.

Next, the strong, compact calibrator 1638+398 was used to determine
the feed polarizations of the antennas using the AIPS task LPCAL.  I
assumed that the VLBA antennas had good quality feeds with relatively
pure polarizations, which allowed me to use a linearized model to fit
the feed polarizations.  Once these were determined, the solutions
were applied to observations of 3C\,279, and to the compact source
0954+658.  The polarization angle calibration of the VLBA was set
using component C4 of 3C\,279 which has been shown to be stable over
many years (Taylor 1998). A small RM of $-$40 \radm2\ was assumed for
C4, consistent with the value of $-$64 $\pm$ 50 \radm2\ found by Taylor
(1998).  The absolute electric vector position angle (EVPA)
calibration is estimated to be good to about 3\deg\ at all frequencies
observed.  This translates into an absolute uncertainty in the RM 
(determined by observations between 8 and 15 GHz) of
50 \radm2.  The polarization calibration of the VLBA was verified by
observing the compact source 0954+658 with the VLA on 1998 August 4.
The flux density scale and polarization angle calibration at the VLA
were set using 3C\,286.  Reasonable agreement was found between the
polarized flux densities and polarization angles measured with the VLA
and VLBA (see Table 3).  The discrepancies are consistent with the
moderately low signal-to-noise ratio of the observations of 0954+658.

\section{Results}

In order to compare images of the total intensity, polarized
intensity, and polarization angle obtained at different frequencies
with a fixed array of antennas, it is necessary to taper the
observations made at higher frequencies (and hence higher intrinsic
resolution) in order to match the resolution of the lowest frequency
observed.  Various combinations of images were produced covering the
ranges from 5-8 GHz, 8-15 GHz, and 12-15 GHz. The heavy taper
required to match the 15 GHz to the 5 GHz resolution made this
combination of limited utility.  And as the cores were 
generally found to depolarize at 5 GHz, all of the results
presented here have been derived from the 7 frequencies in 
the range 8--15 GHz (see Table 2).  Faraday Rotation Measures were obtained
by performing a least-squares fit of a $\lambda^2$-relation
to measurements of the polarization angle at each pixel
in matched-resolution images.  Representative fits are shown
for each source in Figures 1--5, and the RM distributions for 
each target sources is shown in Figure 6.

In the process of fringe-fitting the VLBA data, the brightest
component of each source (in almost all cases the flat-spectrum core
component) is shifted to the map center.  If this core component is
the location in the jet where the synchrotron self absorption optical
depth is unity, then its position will vary with frequency (Blandford
\& K\"onigl 1979), such that higher frequencies look ``deeper'' down
the jet towards the true center of activity.  This frequency dependent
shift of the core component can be measured by using optically thin
features in the jet to register simultaneous, multi-frequency
observations (see e.g. Lobanov 1996).  The amount of this shift is
small, $< 0.2$ mas between 5 and 15 GHz for 3C\,345 (Lobanov 1996).
For all the observations presented here, the registration of the
multi-frequency observations were found to be within 10\% of the beam
size, and no attempt to correct for a core shift was made.

%
%

\subsection{4C\,73.18 (1928+738)}

This relatively bright, nearby quasar ($V$ = 15.5, $z$=0.3021, 1 mas = 5.52
pc -- Lawrence \etal\ 1996) is included in the PR sample (Pearson \&
Readhead 1988), and as such was imaged with VLBI in the early 1980s.
These first VLBI observations were reported on by Eckart \etal\ (1985)
who found a southward jet extending 17 mas with superluminal
components moving at 0.6 mas y$^{-1}$ (14 c) relative to an
assumed stationary core component.  Hummel \etal\ (1992) monitored
1928+738 in six epochs at 22 GHz between 1985.10 and 1989.71 and found
relative superluminal motions in the range 0.3 -- 0.4 mas y$^{-1}$.
Ros \etal\ (1999) were able to align three epochs taken at 5 or 8.4
GHz between 1985.77 and 1991.89 relative to the nearby BL Lac object
2007+777.  Their astrometry suggests that in some epochs the core
component is weak or undetected at 8.4 GHz.

\begin{figure}
\vspace{16.5cm}
\includegraphics{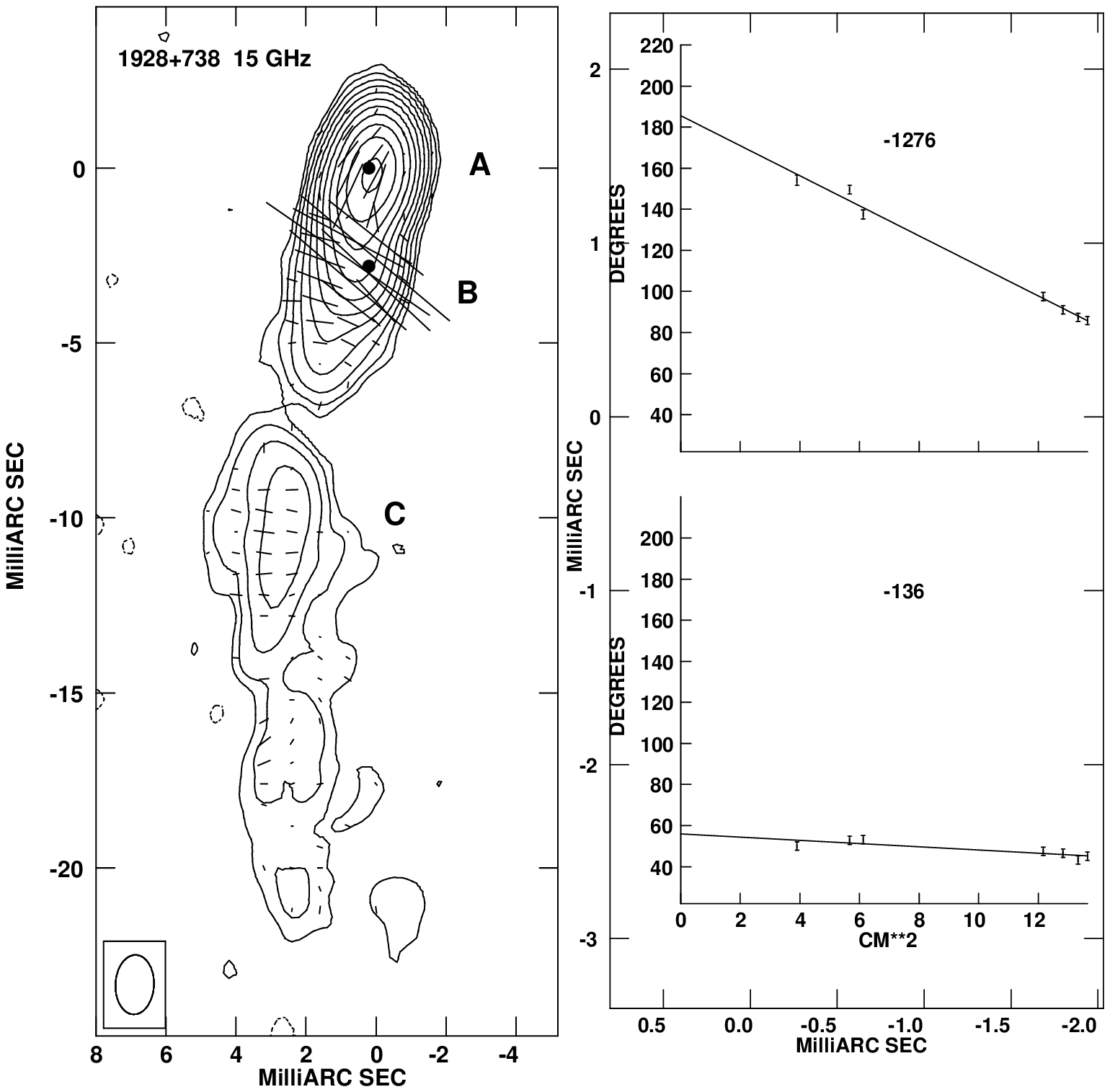}
\figcaption{{\bf (a)} Polarized intensity electric vectors 
(1 mas = 5 mJy/beam; vectors not corrected for Faraday rotation) 
overlaid on total intensity contours for 1928+738 at 15 GHz.  Contours
are plotted at $-$2, 2, 4, ... 2048 mJy/beam with negative
contours shown dashed. The restoring beam is plotted in the lower left
corner and has dimensions 1.7 $\times$ 1.1 mas in position angle
$-$2.6\arcdeg.  The bullets ($\bullet$) indicate representative points
where the rotation measure fit has been plotted.  {\bf (b)} The
electric vector position angle as a function of $\lambda^2$ between
8 and 15 GHz and the
derived RM fits for the points shown in (a).
\label{fig1}}
\end{figure}

\placefigure{fig1}

Cawthorne \etal\ (1993) measured the linear polarization of 1928+738
at 5 GHz (epoch 1984.23) and found a fractional polarization ranging
from a level of 1\% at the northernmost component to 10\% in the knots
$\sim$10 mas to the south.  At both 8 and 15 GHz (epoch 1998.59) I
find 1\% or less fractional polarization from the northernmost
component (A) (see Fig.~1 and Table 4).  The fractional polarization
increases along the jet to 17\% at component C, $\sim$10 mas to the
south.  The RM of component A is high at 2200 \radm2\ in the rest
frame of the source.  However the depolarization of 0.3, indicating
stronger polarization at 8.4 GHz, suggests that frequency dependent
substructure is present which could influence the observed RM.  In the
jet the RMs fall to background levels ($-$32 $\pm$ 50) and are roughly
in agreement with the RM of 33 $\pm$ 4 measured on kiloparsec scales
with the VLA by Wrobel (1993).

\subsection{3C\,395 (1901+319)}

This source was identified with a magnitude 17 quasar at $z$=0.635 (1
mas = 7.75 pc) by Gelderman \& Whittle (1994).  The parsec to
kiloparsec scale radio structure has been reported on by Lara \etal\ (1997).
They find a bright, unresolved core (component A) and a straight jet
extending 15 mas to the east.  While the inner 15 mas of the jet is straight, 
after that it appears to bend by 180\deg\ and form a knotty
jet traveling more than 500 mas in the opposite apparent
direction.  Given the likely orientation of the radio source
close to the line-of-sight, a large intrinsic bend in
the source is not required.  On the arscecond scale, a halo
is seen which Lara \etal\ interpret to to be the radio lobes
seen in projection against the jet.  

\begin{figure}
\vspace{16.5cm}
\includegraphics{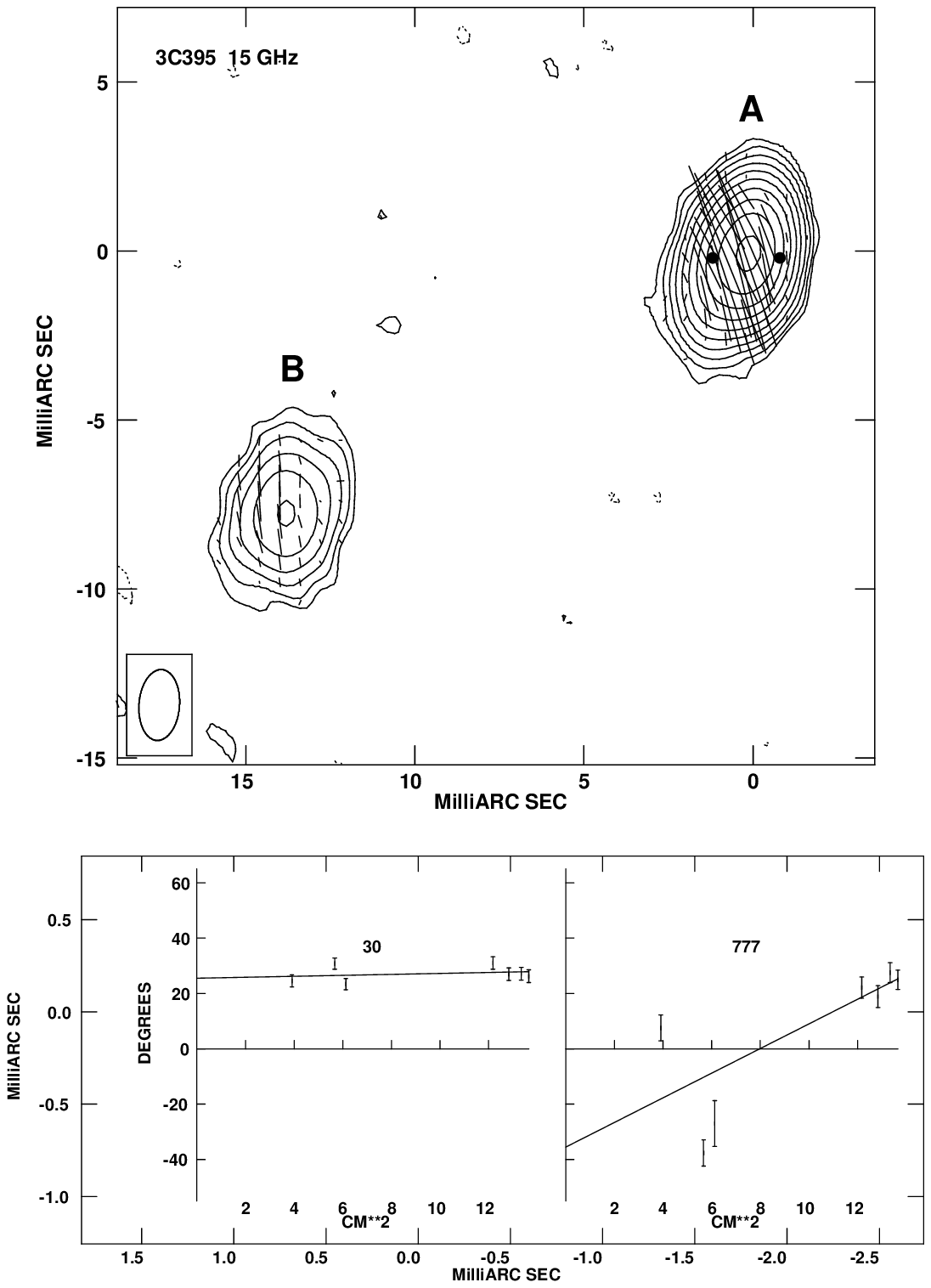}
\figcaption{{\bf (a)} Polarized intensity electric vectors (1 mas =
  3.1 mJy/beam; vectors not corrected for Faraday rotation) 
overlaid on total intensity contours for 3C\,395 at 15 GHz.  Contours
are plotted at $-$1.5, 1.5, 3, ... 768 mJy/beam with negative
contours shown dashed. The restoring beam is plotted in the lower left
corner and has dimensions 2.1 $\times$ 1.2 mas in position angle
$-$4.4\arcdeg.  The bullets ($\bullet$) indicate representative points
where the rotation measure fit has been plotted. {\bf (b)} The
electric vector position angle as a function of $\lambda^2$ between
8 and 15 GHz and the
derived RM fits for the points shown in (a).
\label{fig2}}
\end{figure}

\placefigure{fig2}

No VLBI polarimetric observations of 3C\,395 have been previously
reported.  I find the core and inner jet component (marked with an A
in Fig.~2) to be polarized at the $\sim$1.5\% level on average, and
to have an average RM of 300 \radm2 (see Table 4).  There is a strong
gradient in RM (see Figures 2 and 6), however, such that the western
end reaches RMs of 1200 \radm2, while the eastern end is consistent
with an RM of 0 \radm2. The western end, presumably closest to the
center of activity, has just 1\% polarization at 8.3 GHz.  The 
poor fits of the RM at the western end are probably related to 
the low fractional polarization, and blending with the more 
strongly polarized eastern end.
The 
polarization increases toward the eastern end to a maximum at the edge
of 10\%.  The $\sim$16 mas distant component (B), is $\sim$11\%
polarized at 8.3 GHz and has a small RM of 68 $\pm$ 40 \radm2.  
The integrated RM measured by Simard-Normandin \etal\ (1981) 
was 169 $\pm$ 1 \radm2\ for this moderately low galactic latitude
($b = 12$\deg) source.

\subsection{2134+004}

This magnitude 17 quasar lies at a redshift of 1.94 (1 mas = 8.24 pc)
and has been observed to vary in the optical by more than 3 magnitudes
(Gottlieb \& Liller 1978).  Although variations in total intensity are
only $\sim$15\% on timescales of years (Aller, Aller \& Hughes 1994),
large variations were reported for the parsec-scale structure at 10.7
GHz between 1987 and 1989 requiring apparent velocities of more than
60 c (Pauliny-Toth \etal\ 1990).  These structures are quite
different, however, from the 15 GHz images reported by Kellermann
\etal\ (1998).  Observations presented here support the suggestion by
Kellermann \etal\ that the Pauliny-Toth \etal\ structures were the
result of inadequate sampling of the ($u$,$v$) plane and the close
proximity of this source to the celestial equator.

\begin{figure}
\vspace{16.5cm}
\includegraphics{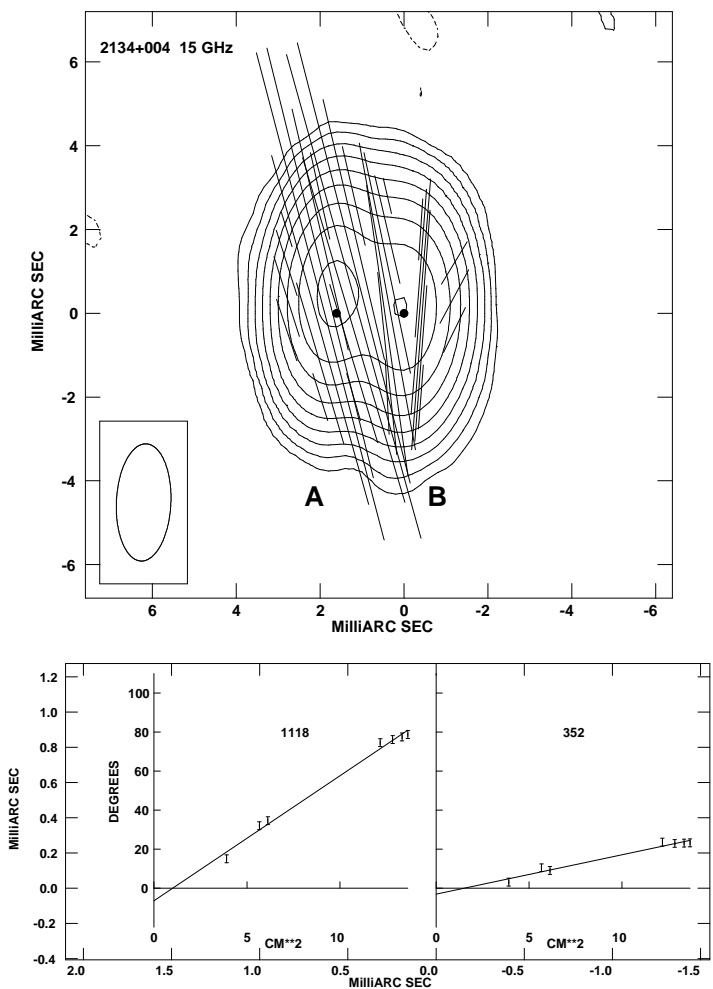}
\figcaption{{\bf (a)} Polarized intensity electric vectors (1 mas = 
12.5 mJy/beam; vectors not corrected for Faraday rotation) 
overlaid on total intensity contours for the inner core and jet
of 2134+004 at 15 GHz.  Contours
are plotted at $-$10, 10, 20, ... 2560 mJy/beam with negative
contours shown dashed. The restoring beam is plotted in the lower left
corner and has dimensions 2.8 $\times$ 1.3 mas in position angle
$-$2.6\arcdeg. 
\label{fig3}}
\end{figure}

\placefigure{fig3}

No VLBI polarimetric observations of 2134+004 have been 
previously reported.  I find two unresolved components
separated by 1.7 mas (see Fig.~3).  Component A is the brighter of the
two at 15 GHz and has a flatter spectral index of 
0.0 $\pm$ 0.1, compared to $-$0.6 $\pm$ 0.1 for component B.  
On the basis of its flatter spectrum I identify
component A as the core component.  Component A is polarized 
at the $\sim$5\% level and exhibits an RM of 1100 \radm2 (see Table 4).
Component B is polarized at the $\sim$3\% level and has a 
significant RM of 340 $\pm$ 20 \radm2.  After correcting 
for the effects of the different RM, both components have a
projected magnetic field direction near 90\deg\ (see Fig. 7),
very nearly parallel to the jet direction.  

Given the strong polarized flux density of component B (96 mJy at 15
GHz), there is some hope that this component may be stable enough to
use for absolute electric vector position angle calibration of VLBI
observations at frequencies of 8 GHz and higher.  Such usage would be
similar to how these and many other VLBI polarimetric observations
have made use of component C4 in 3C\,279 (described in \S3.1).  The
position on the sky of 2134+004 makes it a possible alternate for
observations scheduled in such a way as to make observations of
3C\,279 (1253{\tt -}055) impractical.  One disadvantage of 2134+004 is that
component B is only 1.7 mas (14 pc from the core component (A)), and
still has a considerable observed RM of 340 \radm2 (3000 \radm2\ in
the rest frame of the source).  As an example, an RM of 340
\radm2\ causes a rotation of 17\deg\ between 8.4 and 15 GHz.  Further
observations are also required to check on the stability of the
polarization properties of 2134+004.

\subsection{CTA\,102 (2230+114)}

The highly polarized quasar (HPQ) CTA\,102 lies at a
redshift of 1.037 (1 mas = 8.54 pc).  CTA\,102 has the distinction of
being the first source demonstrated to be variable in the radio
(Scholomitski 1965).  The short timescale ($\sim$100 days) of the
variations implied a small angular size and a high brightness
temperature.  CTA\,102 was also one of the targets of early VLBI
observations by Kellermann \etal\ (1968) who confirmed that the source
flux density was dominated by a compact ($<$4 mas) component.  A
multi-wavelength, global VLBI campaign on CTA\,102 is reported on by
Rantakyr\"o \etal\ (1996), who find a strong compact component,
presumed to be the core, and a jet extending 15 mas to the southeast.
Using their own observations from 1983 to 1992 and those available in
the literature Rantakyr\"o \etal\ find a range of apparent jet
velocities from 0 to 30 $\pm$ 12 c.  These velocities are at
the extreme end of those measured in compact jets (see Vermeulen \&
Cohen 1994), and given the rather coarse sampling of the VLBI
observations, may be the result of misidentifying components.

\begin{figure}
\vspace{16.5cm}
\includegraphics{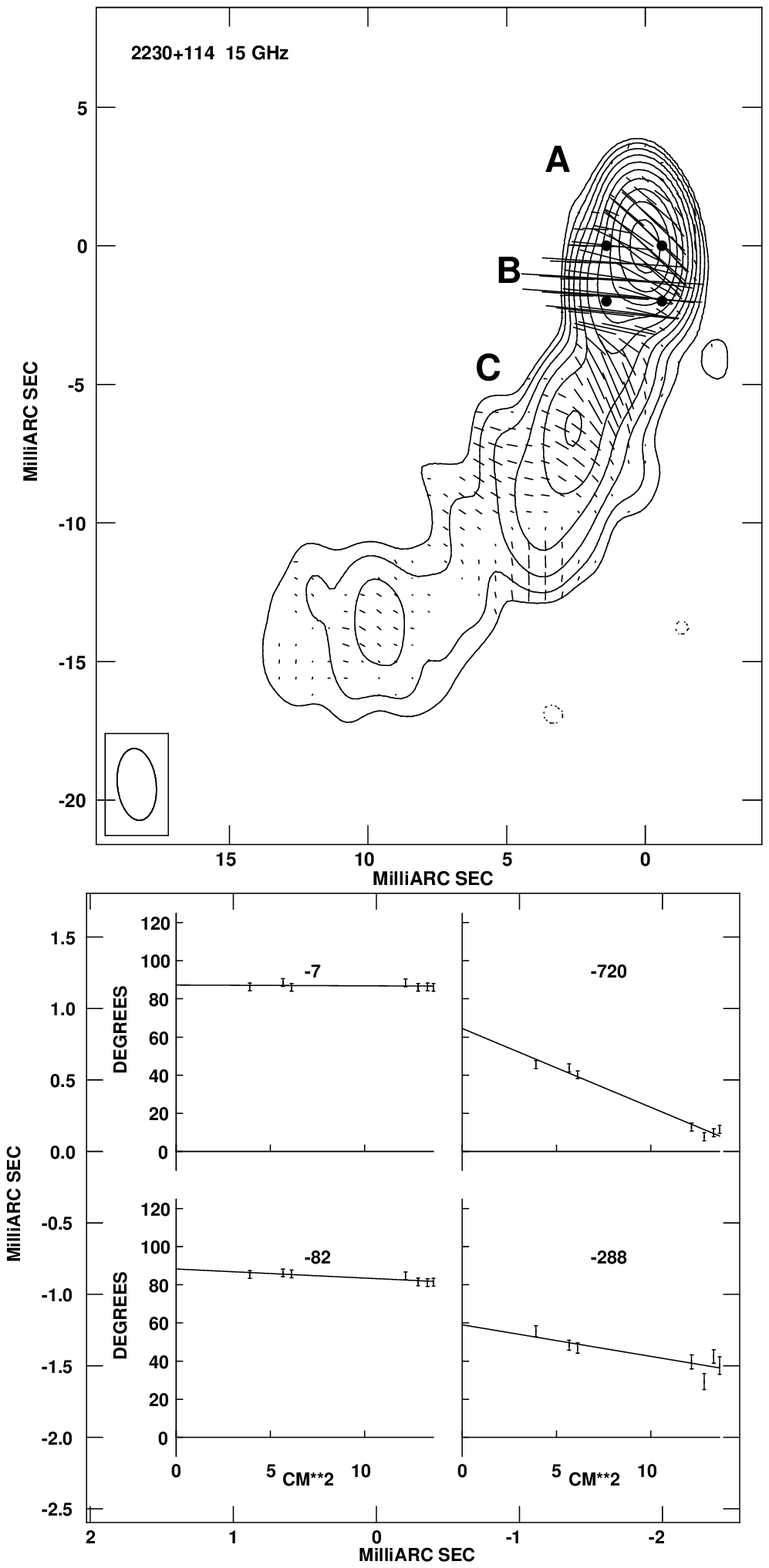}
\figcaption{{\bf (a)} Polarized intensity electric vectors (1 mas = 12.5
  mJy/beam; vectors not corrected for Faraday rotation) 
overlaid on total intensity contours for 2230+114 at 15 GHz.  Contours
are plotted at $-$6, 6, 12, ... 3072 mJy/beam with negative
contours shown dashed. The restoring beam is plotted in the lower left
corner and has dimensions 2.6 $\times$ 1.4 mas in position angle
6\arcdeg.  The bullets ($\bullet$) indicate representative points
where the rotation measure fit has been plotted. {\bf (b)} The
electric vector position angle as a function of $\lambda^2$ between
5 and 8 GHz and the
derived RM fits for the points shown in (a).
\label{fig4}}
\end{figure}

\placefigure{fig4}

No VLBI polarimetric observations of 2230+114 have been previously
reported.  We find a similar source structure to that seen by
Rantakyr\"o \etal\ (1996) at 22 GHz, and by Kellermann \etal\ (1998)
at 15 GHz, consisting of a core (component A), and a wiggling
one-sided jet extending approximately 15 mas to the southeast (Fig.~4).  The
fractional polarization of the core is 1.1\% at 15 GHz, but only 0.5\%
at 8.4 GHz, indicating significant depolarization.  The fractional
polarization of the inner jet component B is much higher, $\sim$13\%
at both 8.4 and 15 GHz.  The jet fades in total intensity with
increasing distance from the core and is polarized at the 5--10\%
level.  The RMs reach $-$1000 \radm2\ in the western part of component
A (Fig.~6).  In component B the polarization angles are well fit by a RM of
$-90\,\pm\,20$ \radm2 (Fig.~4).  The proximity of the more strongly polarized
component B,may have partly obscured the
RM of component A.  The true RM of component A may be considerably
larger. In the jet (e.g., component C) the RMs range
between $\pm$400 \radm2, but are not particularly well fit owing to a
low SNR between 12 and 15 GHz.  This is also reflected in the
anomalously low ratio of $D_{15/8}$ for component C.  On the large
scale, Simard-Normandin \etal\ (1981) measured an RM of
$-53\,\pm\,0.4$ \radm2.

\subsection{Variable rotation measures in 3C\,279}

The strong quasar 3C\,279 was observed briefly in order to use the
stable jet component C4 to set the absolute 
EVPA.  As mentioned in \S3.1, the derived corrections were checked
using contemporaneous VLBA and VLA observations of the compact source
0954+658 and found to be within 13\deg\ at all frequencies.
Comparing these epoch 1998.59 observations to similar
observations made in 1997.07 (Paper I), the properties of C4
appear stable to within $\sim$10\% in total intensity and polarized intensity
and to within 13\deg\ in position angle.  In contrast, at 15 GHz the core
of 3C\,279 increased in total intensity by 14\%, in polarized intensity
by 90\% and decreased in RM by 75\% (see Fig.~5).  To my knowledge,
this is the first observation of variability in the RM of an 
extragalactic radio source.  

\begin{figure}
\vspace{16.5cm}
\includegraphics{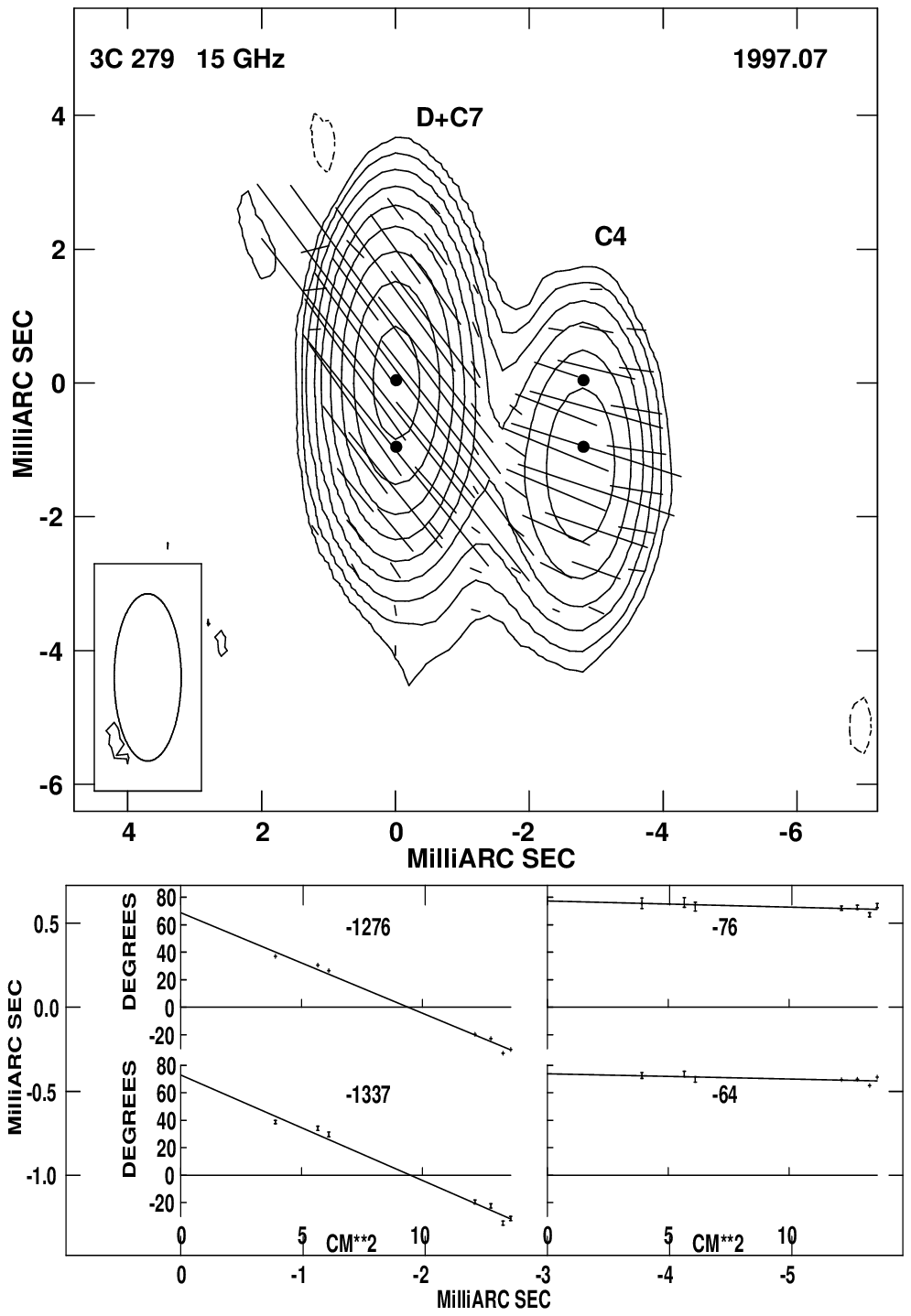}
\includegraphics{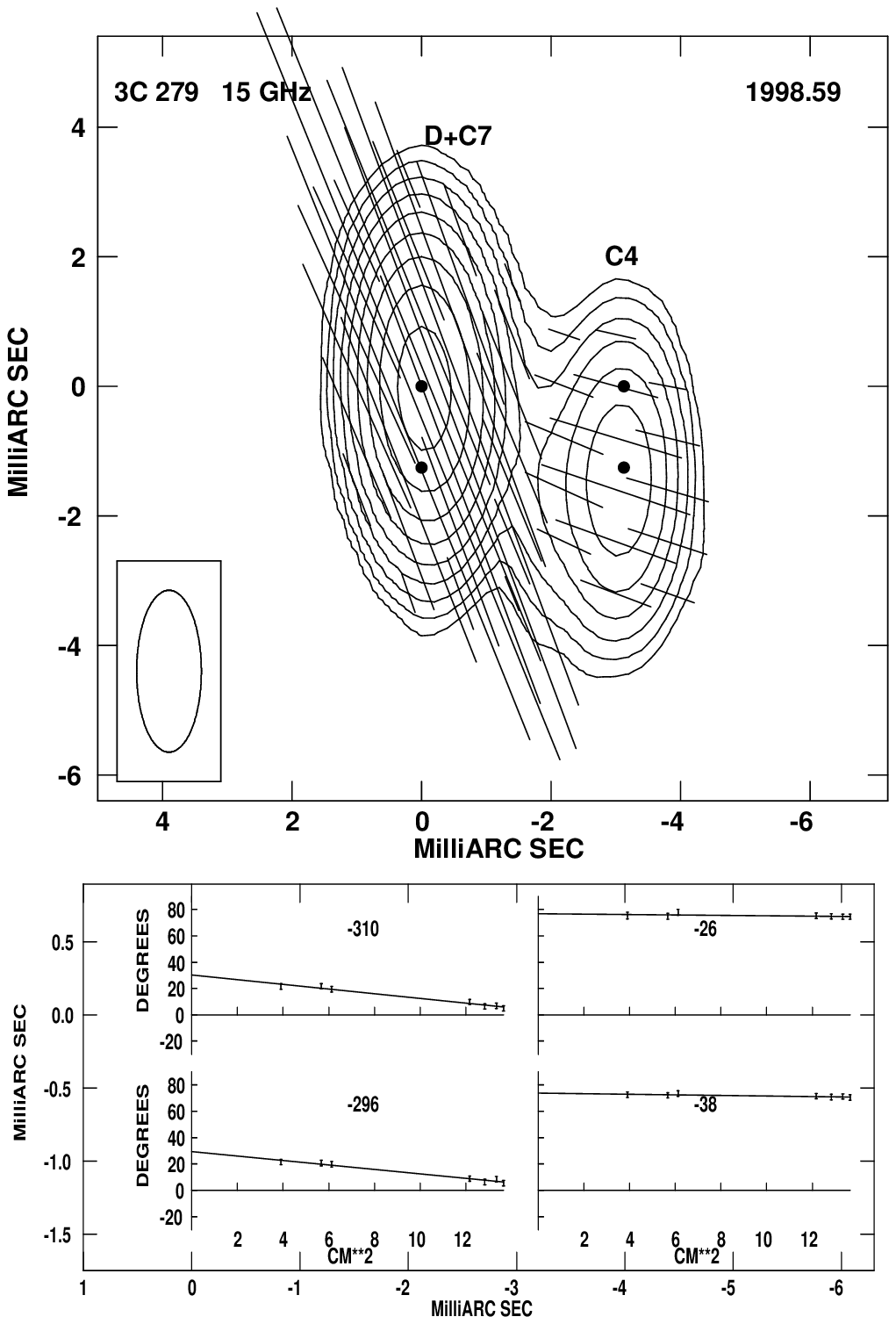}
\figcaption{{\bf (top)} Polarized intensity electric vectors (1 mas =
  100 mJy/beam; vectors not corrected for Faraday rotation)
overlaid on total intensity contours for 3C\,279 at 15 GHz at 
epochs 1997.07 and 1998.59.  Contours
are plotted at $-$50, 50, 100, ... 10240 mJy/beam with negative
contours shown dashed. The restoring beam is plotted in the lower left
corner and has dimensions 2.5 $\times$ 1.0 mas in position angle
0\arcdeg.  The bullets ($\bullet$) indicate representative points
where the rotation measure fit has been plotted. {\bf (bottom)} The
electric vector position angle as a function of $\lambda^2$ between
8 and 15 GHz and the
derived RM fits.      
\label{fig5}}
\end{figure}

\placefigure{fig5}

Taylor (1998) suggested that 3C\,279 could be useful for absolute EVPA
calibration even at frequencies as low as 2 GHz under the assumption
that component C4 dominated the polarized flux density.  Given the
highly variable nature of the core polarization properties, one can
imagine that this assumption may sometimes be invalid, so it is
probably unwise to attempt to tie EVPA calibration to C4 in
observations where C4 cannot be clearly resolved from the core (e.g.
below 5 GHz with the VLBA).  Also the uncertainty of the determination
of the RM based on high frequency observations extrapolates to large
uncertainties in the EVPA of C4 below 8 GHz.

\section{Discussion}

It is possible that the RMs measured for the core and inner jet
components are not produced by a Faraday screen, but have some other
origin (Taylor 1998 and references therein).  One possibility is a
frequency dependent substructure to the inner components.  Certainly
in the cores of many of these sources higher frequency observations
have shown them to be composed of multiple components.  Blending of
nearby components will produce similar effects at the interface where
they merge. It seems unlikely, however, that the changes in EVPA due
to this substructure, or blending of nearby components, would
reproduce a $\lambda^2$-law, especially within the 8 GHz band where
the change in frequency is quite modest (see the clear changes in EVPA
at 8 GHz in 1928+738 shown in Fig.~1).  It is more likely that these
observations sample the RM of the component that dominates the
polarized flux density within the telescope beam.  An analogous, but more extreme, situation is
found in single dish observations of extended sources having high RMs
such as Hydra~A, for which Simard-Normandin \etal\ (1981) find an RM
of $-$871 $\pm$ 8 \radm2\ whereas detailed imaging of the RM structure
by Taylor \& Perley (1993) reveal RMs between $-$12000 and $+$3000
\radm2.  Higher resolution observations are in progress in order to
better resolve the RM structure in a few of the quasars in this
sample.  For the remainder of this discussion we proceed with the
hypothesis that the RMs are the result of a magnetized plasma
somewhere along the line-of-sight.

\begin{figure}
\vspace{19.75cm}
\includegraphics{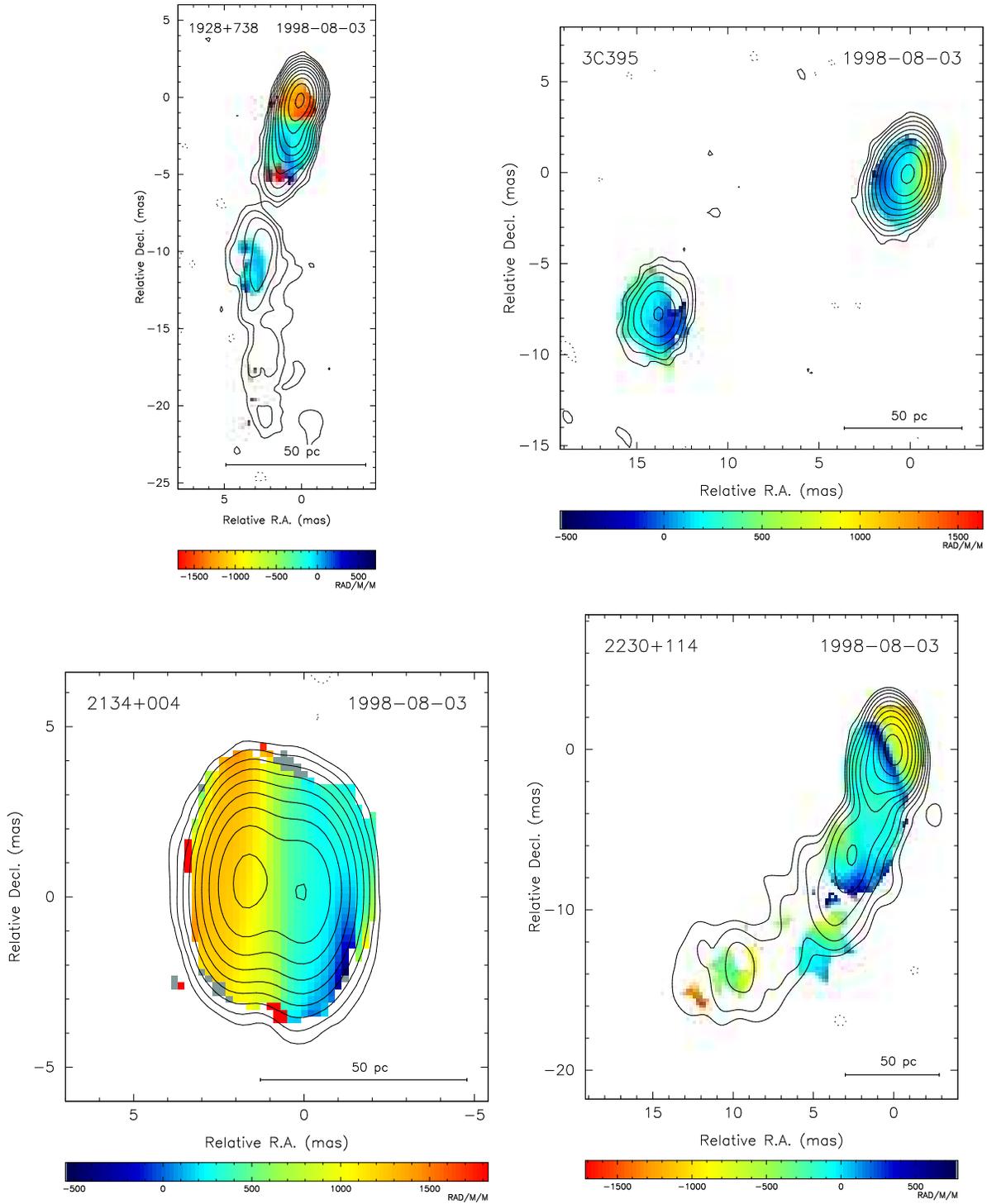}
\figcaption{Rotation measure images of 1928+738, 3C\,395, 2134+004, and 
2230+114 with contours of total intensity superimposed.  Contours are as
in Figures 1, 2, 3, and 4.\label{fig6}}
\end{figure}

\begin{figure}
\vspace{19.75cm}
\includegraphics{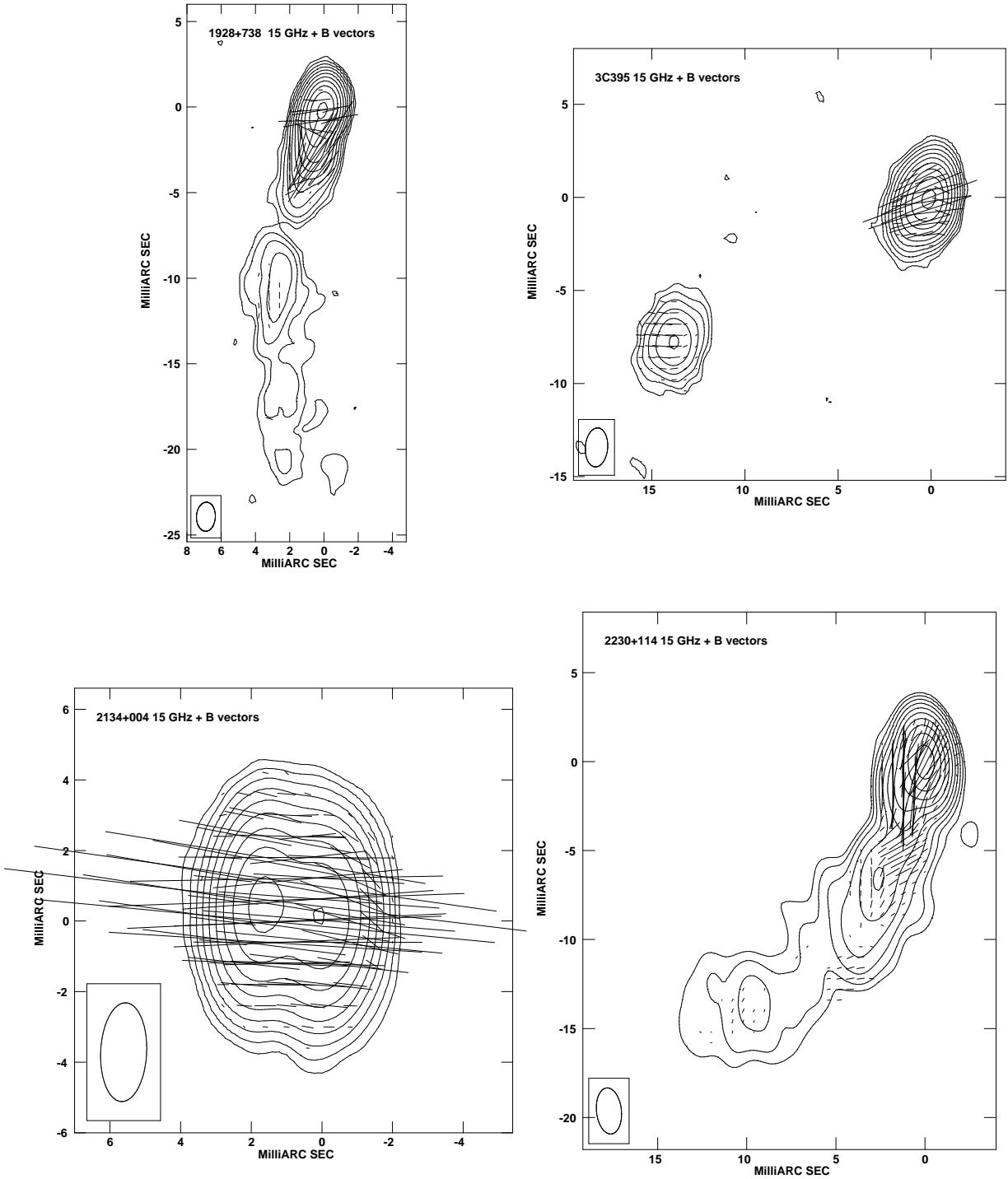}
\figcaption{Projected magnetic field vectors (corrected for the RMs shown
in Fig.~6) for 1928+738, 3C\,395, 2134+004, and 2230+114
with contours of total intensity superimposed.  Contours are as
in Figures 1, 2, 3, and 4.\label{fig7}}
\end{figure}

\placefigure{fig6}
\placefigure{fig7}

Evidence is presented for high Faraday Rotation Measures ($|$RM$| >$
1000 \radm2) near the center of activity in 7 of 8 bright quasars 
(with 3C\,345 being the only exception).  The
jets of 7 of these quasars have $|$RM$|$s less than 100 \radm2
beyond a projected distance from the nucleus of 20 pc --
small enough to be produced by the passage of the radiation through
the ISM of our Galaxy.  For the smallest source, 2134+004, which only
extends over $\sim$14 pc, the RMs are 1120 and 340 \radm2\ 
towards the core and jet component respectively.

\subsection{A cartoon model and implications for the Unified Scheme}

In Fig.~8 a simple cartoon schematic for the AGN environment is shown.
The details of the equatorial region are based on discussions in Peck,
Taylor \& Conway (1999).  According to unified schemes (see for
example the review by Antonucci 1993), core-dominated quasars are
viewed such that the jet axis makes only a small angle to the
line-of-sight, and they therefore exhibit one-sided jets, apparent
superluminal motions, and broad optical emission lines.  While jet
components are within $\sim$100 pc of the center of activity
they are viewed through ionized gas which acts as a Faraday screen.
Once the jet components move farther from the nuclear environment the
RM rapidly drops.

\begin{figure}
\vspace{16.00cm}
\includegraphics{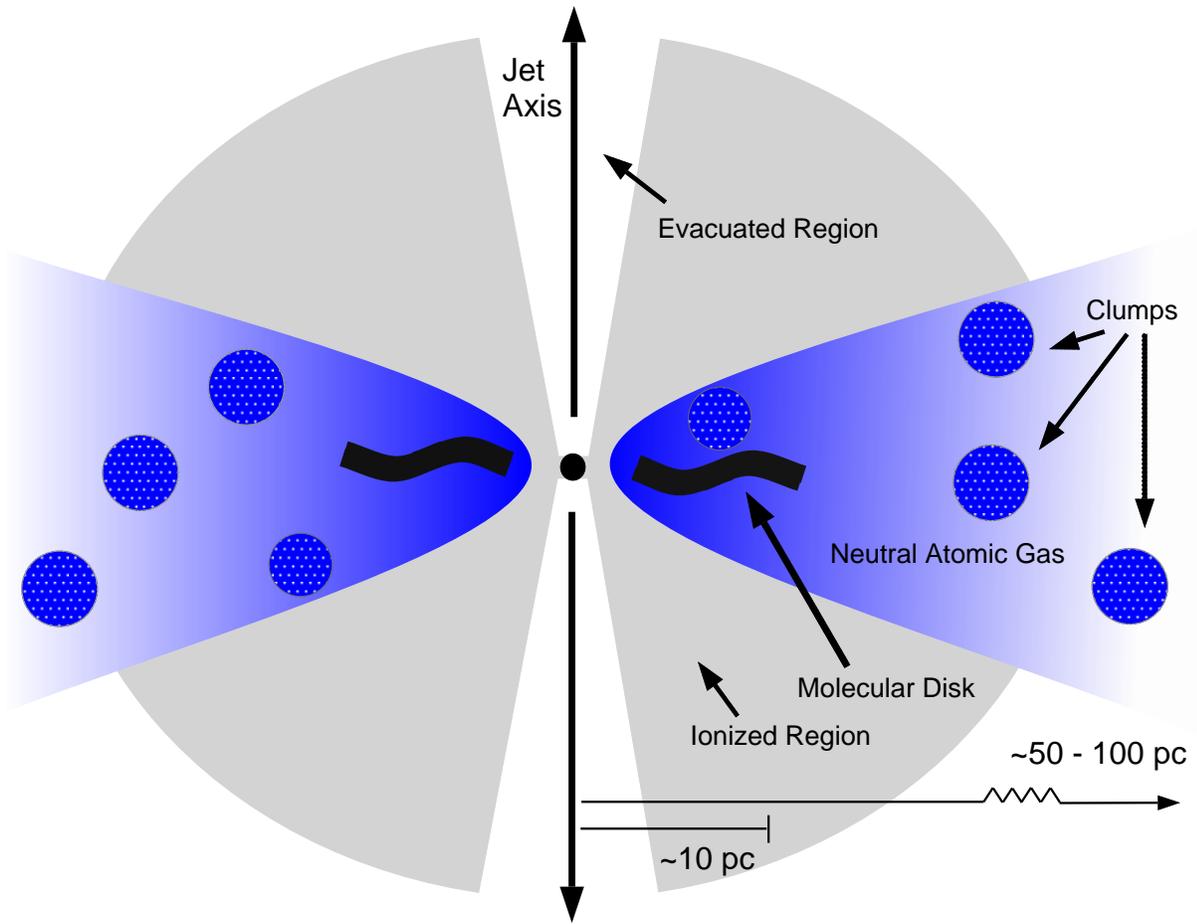}
\figcaption{Cartoon model of the environment around an AGN.  
\label{fig8}}
\end{figure}

\placefigure{fig8}

An observer looking at an AGN nearly edge-on through the denser,
multi-phase disk would likely see a galaxy with narrow optical
emission lines and symmetric parsec-scale radio structures.
The much smaller ($\sim$1 pc) broad line region is hidden from
view by the molecular disk.  While the
cores of lobe-dominated FR II radio galaxies (generally unified with
quasars) have not yet been observed with VLBI polarimetry, there is
another class of bright sources known as Compact Symmetric Objects
that are likely to be youthful versions of FR II radio galaxies
(Readhead \etal\ 1996).  For the CSOs, which often have intrinsic
sizes less than $\sim$100 pc, all the components will be viewed
through a dense multi-phase medium and extremely high RMs ($>>$ 50000
\radm2) are therefore to be expected if the model shown in Fig.~8 is
correct.  Unfortunately such extreme RMs will depolarize the radio
source and hence cannot be measured directly.  Recent VLBA polarimetry
of 20 CSOs by Peck \& Taylor (1999) place typical limits on the
fractional polarization in CSOs of $<$1\%.  No polarized flux on the
parsec-scale has ever been reported for a CSO.

An observer looking at an AGN even closer to the jet axis than the
optimal angle for apparent superluminal motions (1/$\gamma$, where
$\gamma$ is the Lorentz factor of the jet) would see slower motions,
and greater Doppler beaming.  BL Lac objects are even more highly core
dominated than quasars.  There is some evidence that BL Lac objects
have smaller apparent motions than quasars (Gabuzda \etal\ 1994,
Britzen \etal\ 1999).  BL Lac objects have also been observed to have
more strongly polarized cores (Gabuzda \etal\ 1992) compared to quasar
cores.  This might be achieved if the relativistic jet evacuates a
cone through the ionized gas in the nuclear region such that cores of
BL Lacs are not viewed through a Faraday screen (see Fig.~8).  In fact
from lower resolution studies we already know that the BL Lac cores
(which dominate over the jet in integrated polarization) must have
moderate to low RMs (Rusk 1988; Gabuzda \etal\ 1992).  Future studies
of the RM distribution of BL Lac objects on the parsec scale will be
of interest to directly confirm these low RMs.

\begin{figure}
\vspace{16.00cm}
\includegraphics{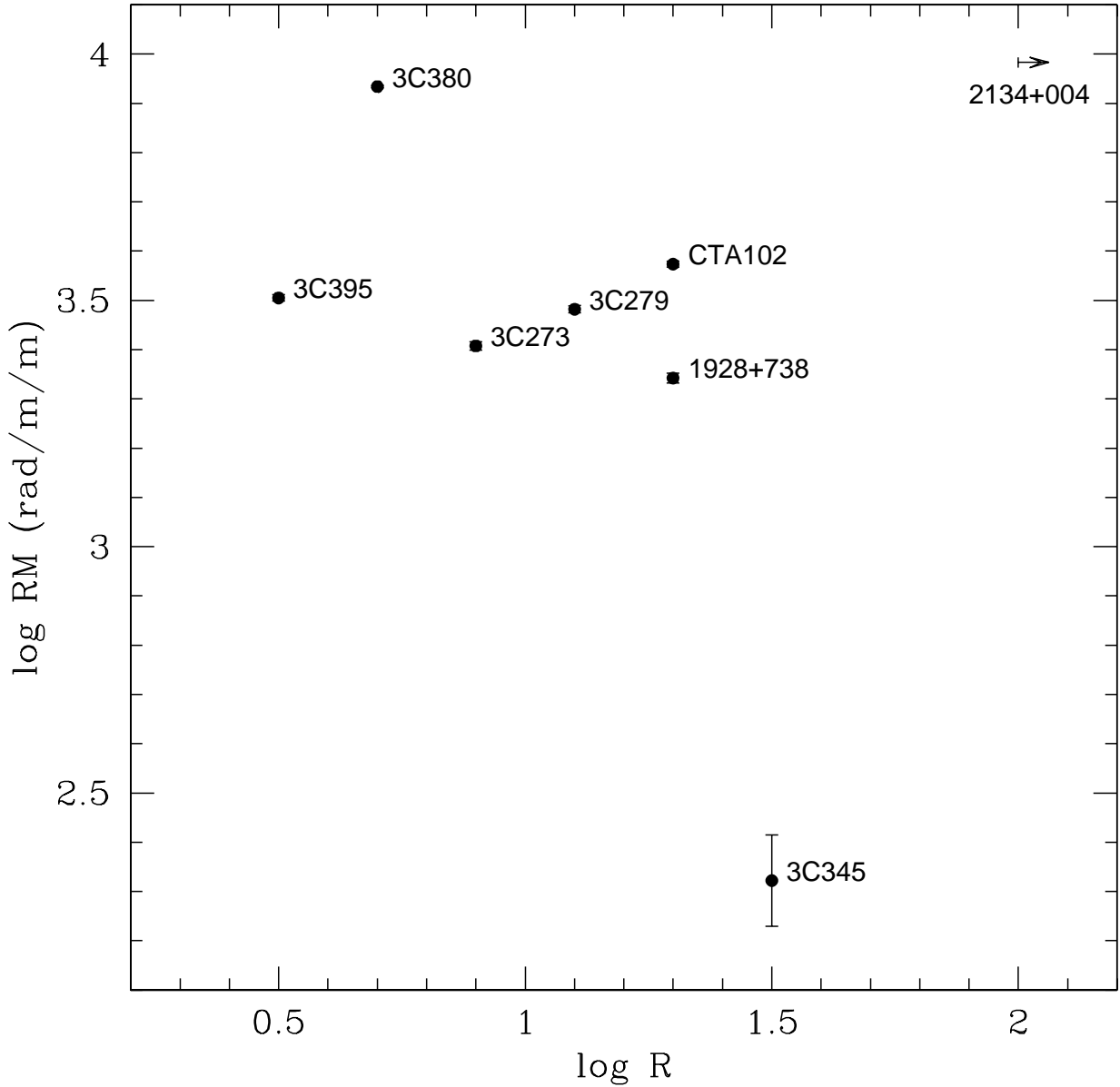}
\figcaption{The maximum RM  observed in the core (corrected by
  (1+z)$^2$) for each source
  plotted against $R$, the ratio of the kpc-scale core flux density to that of
the kpc-scale extended emission at an emitted frequency of 5 GHz.
Values for $R$ have been taken from Vermeulen \& Cohen (1994), except
for
the upper limit on 2134+004 which was computed using the NVSS
image (Condon \etal\ 1998).  
\label{fig9}}
\end{figure}

\placefigure{fig9}

If the RM observed in the core and inner jets is dependent on the
orientation of the source as suggested above, then it might be
expected to correlate with another orientation dependent parameter,
such as $R$ -- the ratio of the core flux density to the lobe flux
density at an emitted frequency of 5 GHz.  Here the term ``core flux
density'' refers to the arcsecond-scale core which includes all the
VLBI components.  Figure 9 shows the RM plotted against $R$ for the 8
quasars studied in this paper and in paper I.  Although 3C\,345 has
both the lowest RM and the highest measured $R$ value, the results for
the sample as a whole are inconclusive.  One oddity is the high RM,
but lack of any kpc-scale extended emission around 2134+004, even
though the core component is not dominant on the parsec-scale like it
is for the rest of the sources shown.  Indeed computing $R$ on the
basis of the parsec-scale emission ($-$0.5) would shift this point to
the top-left corner of the plot.  I suggest that the lack of extended
emission for 2134+004 is due to an intrinsically small size.  If
2134+004 is viewed at a larger angle to the line-of-sight then one
also expects much slower apparent motion in the jet.  Slow motion has
been confirmed in 2134+004 ($<$4 c -- K. Kellermann, private
communication), in sharp contrast to the average speeds of 12 -- 14 c
reported for 3C\,345, 3C\,380 and 3C\,273 (Vermeulen \& Cohen 1994 and
references therein).


%
%

\section{Conclusions}

A sample of 40 strong, compact AGN is identified for a multi-frequency
polarimetric study.  These observations allow imaging of the
Faraday Rotation Measure distribution on the parsec scale.  This is a
relatively new way of probing the AGN environment.  Rest frame Faraday
Rotation Measures in excess of 1000 \radm2\ are found in the nuclear
regions of 7 out of the 8 quasars studied so far.  In all cases the
high RMs are confined to the nuclear region within a projected
distance of 20 pc of the center of activity.  The high RMs
seen in these sources probably originate in the same region that
produces the narrow optical emission lines.  There is some indication
that higher nuclear RMs are seen when an AGN is viewed at larger
angles to the line of sight, and that correspondingly the lowest
nuclear RMs are seen when the line-of-sight is closely aligned to the
jet axis.  These findings are consistent with the predictions of
unified schemes, but the details of the correlation of the nuclear
RM with other orientation-dependent observables are as yet unclear.
Further multi-frequency polarimetric observations are needed for the
remainder of the sample.

\acknowledgments

I am grateful to Alison Peck and Ken Kellermann for many useful
discussions and to Alison Peck again for her help making Fig.~8. 
I thank the referee, John Wardle, for a thorough review including
numerous constructive suggestions.  This
research has made use of data from the University of Michigan Radio
Astronomy Observatory which is supported by the National Science
Foundation and by funds from the University of Michigan.  This
research has made use of the NASA/IPAC Extragalactic Database (NED)
which is operated by the Jet Propulsion Laboratory, Caltech, under
contract with NASA.
 
\clearpage

\renewcommand{\baselinestretch}{1.2}
\normalsize
\begin{center}
TABLE 1 \\
VLBA R{\sc otation} M{\sc easure} I{\sc maging}
\end{center}
\begin{center}
\begin{tabular}{l l c r r r r r r}
\hline
\hline
Source & Name & ID & Mag & z & $S_{15}$ \\
(1) & (2) & (3) & (4) & (5) & (6) \\
\hline
\noalign{\vskip2pt}
0133{\tt +}476 &   DA55 & Q  & 18.0 & 0.86 & 2.22 \\
0202{\tt +}149 &        & G? & 22.1 & -    & 2.29 \\
0212{\tt +}735 &        & BL & 19.0 & 2.37 & 2.69 \\
0336{\tt -}019 &  CTA26 & Q  & 18.4 & 0.85 & 2.23 \\
0355{\tt +}508 &        & EF & -    &  -   & 3.23 \\
0415{\tt +}379 &  3C111 & G  & 18.0 & 0.05 & 5.98 \\
0420{\tt -}014 &        & Q  & 17.8 & 0.92 & 4.20 \\
0430{\tt +}052 &  3C120 & G  & 14.2 & 0.03 & 3.01 \\
0458{\tt -}020 &        & Q  & 18.4 & 2.29 & 2.33 \\
0528{\tt +}134 &        & Q  & 20.0 & 2.06 & 7.95 \\
0552{\tt +}398 &  DA193 & Q  & 18.0 & 2.37 & 5.02 \\
0605{\tt -}085 &        & Q  & 18.5 & 0.87 & 2.80 \\
0736{\tt +}017 &        & Q  & 16.5 & 0.19 & 2.58 \\
0748{\tt +}126 &        & Q  & 17.8 & 0.89 & 3.25 \\
0923{\tt +}392 & 4C39.25 & Q & 17.9 & 0.70 & 10.84 \\
1055{\tt +}018 &        & BL & 18.3 & 0.89 & 2.15 \\
1226{\tt +}023 & 3C273  & Q & 12.9 & 0.16 & 25.72 \\
1228{\tt +}126 &  M87 & G & 9.6 & 0.00 & 2.40 \\
1253{\tt -}055 & 3C279  & Q & 17.8 & 0.54 & 21.56 \\
1308{\tt +}326 &        & BL & 19.0 & 1.00 & 3.31 \\
\end{tabular}
\end{center}
\clearpage
\begin{center}
{\sc Table 1 Continued}\\
\smallskip
\begin{tabular}{l l c r r r r r r}
\hline
\hline
Source & Name & ID & Mag & z & $S_{15}$ \\
(1) & (2) & (3) & (4) & (5) & (6) \\
\hline
\noalign{\vskip2pt}
1546{\tt +}027 &        & Q & 18.0 & 0.41 & 2.83 \\
1548{\tt +}056 &        & Q & 17.7 & 1.42 & 2.83 \\
1611{\tt +}343 &  DA406 & Q & 17.5 & 1.40 & 4.05 \\
1641{\tt +}399 & 3C345 & Q & 16.0 & 0.59 & 8.48 \\
1741{\tt -}038 &       & Q & 18.6 & 1.05 & 4.06 \\
1749{\tt +}096 &       & BL & 16.8 & 0.32 & 5.58 \\
1803{\tt +}784 &       & BL & 17.0 & 0.68 & 2.05 \\
1823{\tt +}568 &       & BL & 18.4 & 0.66 & 2.31 \\
1828{\tt +}487$^1$ & 3C380 & Q & 16.8 & 0.69 & 1.82 \\
1901{\tt +}319 & 3C395 & Q & 17.5 & 0.64 & 1.09 \\
1928{\tt +}738 &       & Q & 16.5 & 0.30 & 3.04 \\
2005{\tt +}403 &       & Q & 19.5 & 1.74 & 2.51 \\
2021{\tt +}317 &       & EF & - & - &  2.02 \\
2021{\tt +}614 &       & G & 19.5 & 0.23 & 2.21 \\
2134{\tt +}004 &       & Q & 16.8 & 1.93 & 5.51 \\
2200{\tt +}420 & BL Lac & BL & 14.5 & 0.07 & 3.23 \\
2201{\tt +}315 &       & Q & 15.5 & 0.30 & 3.10 \\
2223{\tt -}052 & 3C446 & BL & 17.2 & 1.40 & 3.92 \\
2230{\tt +}114 & CTA102 & Q & 17.3 & 1.04 & 2.33 \\
2251{\tt +}158 & 3C454.3 & Q & 16.1 & 0.86 & 8.86 \\
\hline
\end{tabular}\\
\medskip
{\sc Notes to Table 1}\\
\end{center}
\vspace{-0.1cm}
$^1$ Not in the Kellermann et al.\ (1998) survey.
Col.(1).---B1950 source name.
Col.(2).---Alternate common name.
Col.(3).---Optical magnitude.
Col.(4).---Optical identification from the literature (NED). Key 
to identifications: Q--quasar; G--galaxy; BL--BL Lac object; EF--empty 
field.
Col.(5).---Redshift.
Col.(6).---Total flux density at 15 GHz measured by Kellermann \etal\
(1998), or in the case of 3C\,380 by Taylor (1998).
\smallskip
\clearpage

\begin{center}
TABLE 2 \\
\smallskip
VLBA O{\sc bservational} P{\sc arameters}
\smallskip
 
\begin{tabular}{l r r r r r r r r}
\hline
\hline
Source & Frequency & BW & Scans & Time \\
(1) & (2) & (3) & (4) & (5) \\
\hline
\noalign{\vskip2pt}
0954+658 & 4.616, 4.654, 4.854, 5.096 & 8 & 3 & 10 \\ 
         & 8.114, 8.209, 8.369, 8.594 & 8 & 3 & 10 \\
         & 12.115, 12.591 & 16 & 2 & 10 \\
         & 15.165 & 32 & 2 & 10 \\
3C\,279 & 4.616, 4.654, 4.854, 5.096 & 8 & 2 & 8 \\ 
       & 8.114, 8.209, 8.369, 8.594 & 8 & 2 & 7 \\
       & 12.115, 12.591 & 16 & 2 & 7 \\
       & 15.165 & 32 & 2 & 7 \\
1638+398 & 4.616, 4.654, 4.854, 5.096 & 8 & 7 & 25 \\ 
         & 8.114, 8.209, 8.369, 8.594 & 8 & 7 & 25 \\
         & 12.115, 12.591 & 16 & 7 & 25 \\
         & 15.165 & 32 & 7 & 25 \\
1928+738 & 4.616, 4.654, 4.854, 5.096 & 8 & 8 & 28 \\ 
       & 8.114, 8.209, 8.369, 8.594 & 8 & 8 & 28 \\
       & 12.115, 12.591 & 16 & 8 & 28 \\
       & 15.165 & 32 & 8 & 28 \\
3C\,395 & 4.616, 4.654, 4.854, 5.096 & 8 & 7 & 25 \\ 
       & 8.114, 8.209, 8.369, 8.594 & 8 & 7 & 25 \\
       & 12.115, 12.591 & 16 & 7 & 25 \\
       & 15.165 & 32 & 7 & 25 \\
2134+004 & 4.616, 4.654, 4.854, 5.096 & 8 & 7 & 25 \\ 
       & 8.114, 8.209, 8.369, 8.594 & 8 & 7 & 25 \\
       & 12.115, 12.591 & 16 & 7 & 25 \\
       & 15.165 & 32 & 7 & 25 \\
\end{tabular}
\end{center}
\smallskip
\clearpage

\begin{center}
TABLE 2 Continued\\
\smallskip
\begin{tabular}{l r r r r r r r r}
\hline
\hline
Source & Frequency & BW & Scans & Time \\
(1) & (2) & (3) & (4) & (5) \\
\hline
\noalign{\vskip2pt}
2230+114 & 4.616, 4.654, 4.854, 5.096 & 8 & 7 & 25 \\ 
       & 8.114, 8.209, 8.369, 8.594 & 8 & 7 & 25 \\
       & 12.115, 12.591 & 16 & 7 & 25 \\
       & 15.165 & 32 & 7 & 25 \\
\hline
\end{tabular}
\end{center}
\begin{center}
{\sc Notes to Table 2}
\end{center}
Col.(1).---Source name.
Col.(2).---Observing frequency in GHz.
Col.(3).---Total spanned bandwidth in MHz.
Col.(4).---Number of scans (each 2 -- 4 minutes duration).
Col.(5).---Total integration time on source in minutes.
\bigskip
 
\begin{center}
\clearpage 
TABLE 3 \\
\smallskip
P{\sc olarization} A{\sc ngle} C{\sc alibration}
\smallskip
 
\begin{tabular}{l r r r r r r r r r}
\hline
\hline
Source & $\nu$ & $S_{\rm VLA}$ & $S_{\rm VLBA}$ & $P_{\rm VLA}$ &
$P_{\rm VLBA}$ & $\chi_{\rm VLA}$ & $\chi_{\rm VLBA}$ \\
(1) & (2) & (3) & (4) & (5) & (6) & (7) & (8) \\
\hline
\noalign{\vskip2pt}
0954+658 & 4.7 & 0.33 & 0.29 & 13 & 8.4 & $-$12 & $-$25 \\
         & 8.4 & 0.34 & 0.34 & 10 & 5.5 & $-$22 & $-$29 \\
         & 15.2 & 0.32 & 0.32 & 13 & 17 & $-$15 & $-$28 \\
\hline
\end{tabular}
\end{center}
\smallskip
\begin{center}
{\sc Notes to Table 3}
\end{center}
Col.(1).---Source name.
Col.(2).---Observing band in GHz.
Col.(3).---Integrated VLA flux density in Jy.
Col.(4).---Integrated VLBA flux density in Jy.
Col.(5).---Integrated VLBA polarized flux density in mJy.
Col.(6).---Integrated VLA polarized flux density in mJy.
Col.(7).---VLA polarization angle (E-vector) in degrees.
Col.(8).---VLBA peak polarization angle (E-vector) in degrees.
\bigskip
 
\begin{center}
\clearpage 
TABLE 4 \\
\smallskip
C{\sc omponent} P{\sc ositions,} F{\sc lux} D{\sc ensities, and},
R{\sc otation} M{\sc easures}
\smallskip
 
\begin{tabular}{l r r r r r r r r r r r r r }
\hline
\hline
Comp & $r$ & $I_{\rm 15}$ & $P_{\rm 15}$ & $m_{\rm 15}$ & $\chi_{\rm 15}$ &
       $I_{\rm 8}$ & $P_{\rm 8}$ & $m_{\rm 8}$ & $\chi_{\rm 8}$ &
        RM & BPA & $D_{\rm 15/8}$ \\
(1) & (2) & (3) & (4) & (5) &
      (6) & (7) & (8) & (9) & 
      (10) & (11) & (12) & (13) \\
\hline
\noalign{\vskip2pt}
1928+738 \\
A & 0.0 & 2310 & 6.0 & 0.3 & $-$29 & 1970 & 19 & 1.0 & $-$86 & $-$1300 & 78 & 0.3 \\
B & 2.6 & 420 & 36.3 & 8.6 & 51 & 480 & 44 & 9.2 & 48 & $-$140 & 32 & 0.9 \\
C & 11.1 & 17 & 3.0 & 17 & 79 & 24 & 4.0 & 17 & 85 & $-$32 & 6 & 1.0 \\
3C\,395 \\
A & 0.0 & 890 & 13.0 & 1.5 & 19 & 834 & 13.2 & 1.6 & 30 & 300 & $-$81 & 0.9 \\
B & 15.9 & 51 & 2.6 & 5.1 & 4.9 & 84 & 9.6 & 11.4 & 7.0 & 68 & $-$89 & 0.5 \\
2134+004 \\
A & 0.0 & 3170 & 183 & 5.8 & 14.5 & 3240 & 140 & 4.3 & 77 & 1120 & 84 & 1.3 \\
B & 1.7 & 2610 &  96 & 3.7 &  3.5 & 3810 & 103 & 2.7 & 24 & 340 & 89 & 1.4 \\
2230+114 \\
A & 0.0 & 4490 & 48.1 & 1.1 & 49    & 2740 & 13.1 & 0.5 & 15 & $-$610  & $-$26 & 2.2 \\
B & 2.3 & 580 & 75.6 & 13.0 & 85    & 710 & 97.6 & 13.7 & 82 & $-$90 & 0 & 0.9 \\
C & 7.4 & 200 & 7.4 &  3.7 & 61     & 310 & 20.6 & 6.6 & 75 & $-$185 & $-$1 & 0.6 \\
\hline
\end{tabular}
\end{center}
\smallskip
\begin{center}
{\sc Notes to Table 4}
\end{center}
Col.(1).---Component name.
Col.(2).---Distance from the core in mas.
Col.(3).---15.165 GHz total intensity in mJy/beam.
Col.(4).---15.165 GHz polarized intensity in mJy/beam.
Col.(5).---15.165 GHz fractional polarization in \%.
Col.(6).---15.165 GHz electric vector polarization angle in degrees.
Col.(7).---8.369 GHz total intensity in mJy/beam.
Col.(8).---8.369 GHz polarized intensity in mJy/beam.
Col.(9).---8.369 GHz fractional polarization in \%.
Col.(10).---8.369 GHz electric vector polarization angle in degrees.
Col.(11).---Observed Faraday Rotation Measure in rad m$^{-2}$ derived
from a linear least-squares fit to the polarization angle measurements
between 8.114 and 15.165 GHz as discussed in the text.  If generated
in the rest frame of the source these values are larger by a factor of
$(1+z)^2$.
Col.(12).---Magnetic Field polarization angle corrected for RM.
Col.(13).---Depolarization ($m_{15}/m_{8}$).
\bigskip

\renewcommand{\baselinestretch}{1.5}

\clearpage

%
%


\vfill

\vfill

\clearpage
\end{document}